\documentclass[10pt]{article}
\usepackage{amsmath}
\usepackage{amssymb}
\usepackage{graphicx}

\def\jgr{J. Geophys. Res. }

\begin{document}

\setcounter{figure}{0}
\setcounter{table}{0}
\setcounter{footnote}{0}
\setcounter{equation}{0}

\vspace*{0.5cm}

\noindent {\Large COMPARISON OF CPO AND FCN EMPIRICAL MODELS}
\vspace*{0.7cm}

\noindent\hspace*{1.5cm} Z.M. MALKIN\\
\noindent\hspace*{1.5cm} Central Astronomical Observatory at Pulkovo of RAS\\
\noindent\hspace*{1.5cm} Pulkovskoe Ch. 65, St. Petersburg 196140, Russia\\
\noindent\hspace*{1.5cm} e-mail: malkin@gao.spb.ru\\

\vspace*{0.5cm}

\noindent {\large ABSTRACT.}
In this presentation, several publicly available empiric models of the
celestial pole offset (CPO) and free core nutation (FCN), included those
developed by the author, are investigated and compared each other
from different points of view, such as representation of the observation
data, FCN parameters variation, prediction accuracy. Based on this study,
some practical recommendations are proposed.

\vspace*{1cm}

\noindent {\large 1. INTRODUCTION}

\begin{figure}[ht]
\centering
\includegraphics[clip,width=\textwidth]{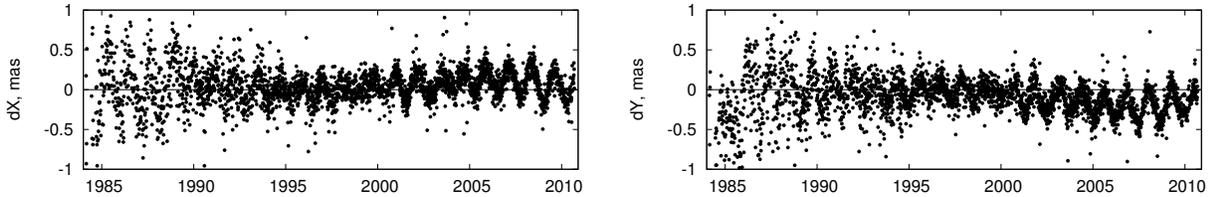}
\caption{IVS combined CPO solution.}
\label{fig:cpo}
\end{figure}

\smallskip
The motion of the celestial pole (CP) is one of the principal constituents of the Earth rotation, and its accurate modelling is needed for many applications.
However, even the most accurate modern theory IAU2003/2006 provides the accuracy of the CP position at a level of about 0.5~mas, which is not sufficient
for many practical tasks.
CPO includes free core nutation (FCN) oscillation with period of about 430 days, other periodic and quasi-periodic terms, and trends.
This complicated CPO structure can be clearly seen in Fig~\ref{fig:cpo}, where the differences between the CP coordinates observed by very long baseline
interferometry (VLBI) and the IAU model are shown.
For VLBI, the IVS (International VLBI Service for Geodesy and Astrometry) combined solution was used (B\"ockmann et al. 2010).

The FCN is often thought to be the largest component of the CPO.
However, one can see that both FCN signal and sum of other CPO components are of similar value of about 0.2--0.3~mas in our days.
Thus the FCN is only a part of the CPO, even not prevailing, but sometimes they are mixed in practical use, e.g. it seems to be the case for the IERS
Conventions (McCarthy \& Petit 2004) that recommends using an FCN model for accurate coordinate transformation between celestial and terrestrial
reference systems.

It seems reasonable and practical to use FCN models for geophysical studies, such as core-mantle coupling or inner Earth's rheological structure, and
use CPO models for applications related to astronomy, in particular for coordinate transformation between terrestrial and celestial systems.

\vspace*{1cm}

\noindent {\large 2. CPO AND FCN MODELS COMPARISON}

\smallskip

In practice, three CPO and two FCN models are available for user's choice:

--- OPA/IERS C04 CPO series (IERS AR 2007, ftp://hpiers.obspm.fr/iers/eop/eopc04\_05/)

--- NEOS/IERS CPO series (IERS AR 2007, ftp://maia.usno.navy.mil/ser7/)

--- Pulkovo ZM2 CPO series (Malkin 2007, http://www.gao.spb.ru/english/as/persac/index.htm)

--- OPA/IERS SL FCN series (Lambert 2009, http://syrte.obspm.fr/~lambert/fcn/)

--- Pulkovo ZM1 FCN series (Malkin 2007, http://www.gao.spb.ru/english/as/persac/index.htm)

In fact, only the SL model has a parametric representation, others are given in the form of $dX$ and $dY$ time series.
However, we refer to them also as models, non-parametric though, because they are a result of data processing as well.

Also, two other FCN models are known: MHB (Herring et al. 2002) and GVS (Gubanov 2010).
However, the MHB model is not supported anymore, and, in fact, is replaced by the SL model constructed in a way close to MHB.
The GVS model is not publicly available in parametric or time series form either.
So, these models were not used in this study.

All the CPO and FCN models are results of approximation of CPO time series observed by VLBI.
IERS C04 CPO solution is a combination of almost all VLBI series available including both individual solutions from IVS Analysis Centers
and IVS combined solution (IERS AR 2007).
In turn, C04 series is used to evaluate SL model (Lambert 2009).
NEOS CPO series is constructed using several selected VLBI series (IERS AR 2007).
ZM2 series is constructed using combining IVS CPO series (Malkin 2007).

\begin{figure}[ht]
\centering
\includegraphics[clip,width=\textwidth]{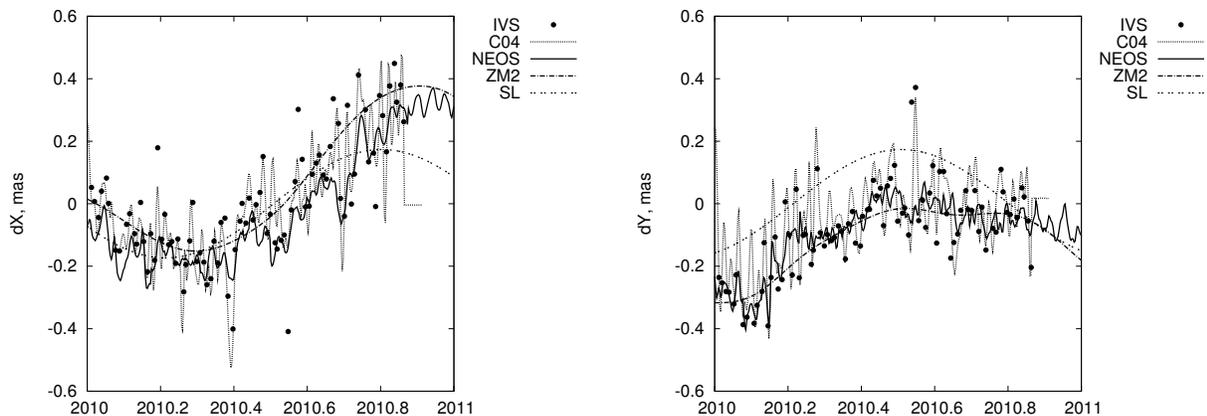}
\caption{IVS combined CPO solutions and models.}
\label{fig:models}
\end{figure}

One of the most important application of CPO in space geodesy is its use in coordinate transformation between terrestrial and celestial frames.
In principle, only full CPO models C04, NEOS and ZM2 should be compared from this point of view.
However, we also include SL model in this analysis because in is recommended by IERS as a replacement for CPO (McCarthy \& Petit 2004).
Figure~\ref{fig:models} shows the representation of the actual CP motion as observed by VLBI and all four models for 2010.

Some features  of compared models are clearly visible in the plots.
SL model shows substantial bias wit respect to VLBI data and other models, which directly follows from the method of construction of this model.
C04 models evidently includes high frequency components, may be not corresponding to the actual CP movement; this model
also does not include prediction, which makes is unsuitable for near real-time data processing.
NEOS model is much closer to IVS series but sometimes is biased, which can be explained by using of different time series.
ZM2 shows the best representation of the IVS combined solution, which we consider as the IERS standard for $dX$ and $dY$ data.

\begin{figure}[ht]
\centering
\includegraphics[clip,width=\textwidth]{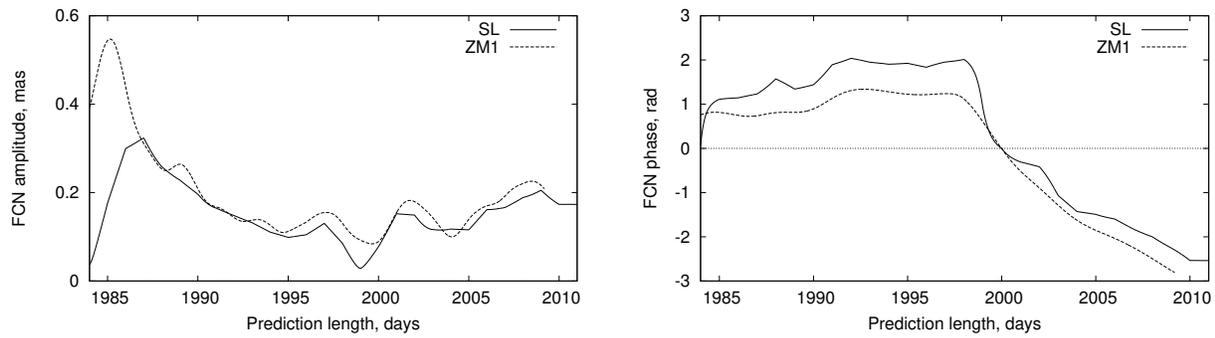}
\caption{FCN amplitude and phase variations.}
\label{fig:fcn_par}
\end{figure}

As mentioned above, FCN observations and modelling allows us to gain knowledge of properties of the inner Earth.
In particular, important conclusions can be derived from the FCN amplitude and phase variations.
As was shown previously (Malkin 2007) MHB, SL, and ZM1 models show similar behavior of the FCN amplitude and phase.
Updated results for SL and ZM1 models are depicted in Fig~\ref{fig:fcn_par}.
One can see from this comparison that both models reveal very similar variations in the FCN amplitude and phase,
except the amplitude in the beginning of the interval, where VLBI results have relatively low accuracy as it can be seen in Fig.~\ref{fig:cpo}.
As to the phase variations, SL model provides better time resolution and cover longer time interval including prediction.
Similar results were obtained for the GVS model (Gubanov 2010).
It can be mentioned here that the method used for construction of the ZM1 model allows one to compute the FCN series with any desirable
time resolution just adjusting the wavelet parameter (Malkin 2007).
Detailed discussion on physical reasons of observed phase and amplitude changes is beyond of the scope of this paper.
For example, one of interesting observation is correlation between the FCN phase jumps and geomagnetic jerks (Shirai et al. 2005)

\begin{figure}[ht]
\centering
\includegraphics[clip,width=\textwidth]{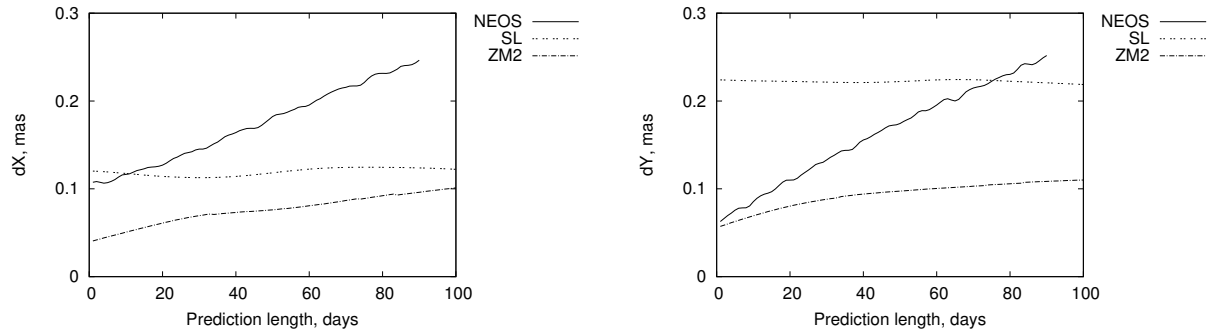}
\caption{CPO prediction errors, see (Malkin 2010) for details.}
\label{fig:pred}
\end{figure}

Since VLBI CPO results are available with delay 1-2~weeks for rapid EOP sessions, CPO prediction is needed for applications requiring
real-time and forecast CPO values.
Prediction accuracy of different CPO models was assessed in (Malkin 2010), and result is shown in Fig.~\ref{fig:pred}

\begin{figure}[ht]
\centering
\includegraphics[clip,width=0.9\textwidth]{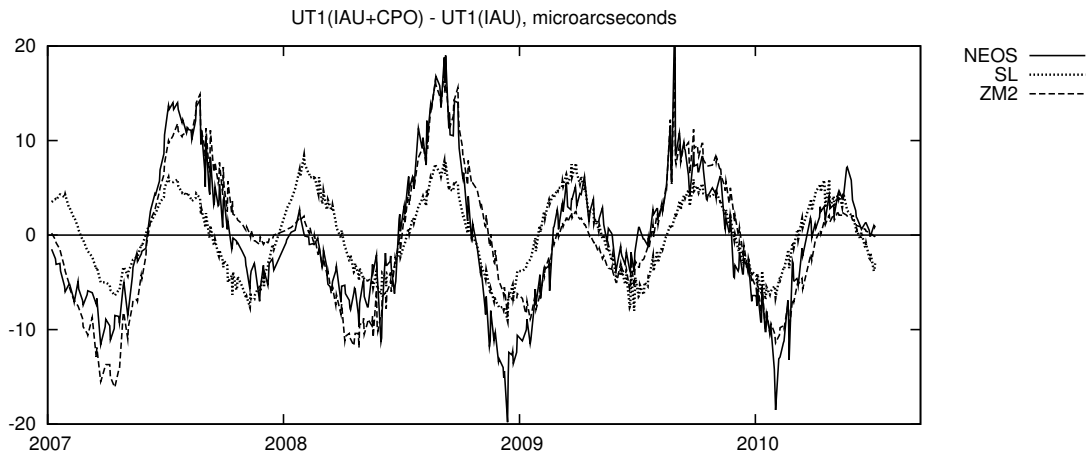}
\caption{Results of UT1 Intensives processing with different CPO models.}
\label{fig:ut1_int}
\end{figure}

A practical impact of different CPO modelling on UT1 estimates obtained from VLBI Intensives series.
These observations are performed on restricted network of 2-3 stations during short session duration
In Fig.~\ref{fig:ut1_int}, differences are shown between UT1 estimates obtained with given CPO model and without CPO modelling (zero model).
As shown above, in our days, FCN contribution to CPO is at a level of 50-60\%.
In correspondence with this fact, UT1 estimates obtained with SL (FCN) model lie between UT1 estimates obtained with NEOS and ZM2 (CPO) and with zero
models.

\clearpage
%\vspace*{1cm}

\noindent {\large 4. CONCLUSIONS}

\smallskip

Modelling of the CPO is necessary for highly accurate applications in astronomy, geodesy, navigation and other fields, particularly required
transformation between terrestrial and celestial coordinate systems with accuracy better than several tenths of milliarcsecond.
Currently, several CPO and FCN models are available, and all of them are used by different users.
Some of them are the FCN models, which represent only a part of the CPO impact on the CP motion.
These models have been compared in this work.

The FCN models are most suitable for geophysical studies.
Two models compared here, ZM1 developed in the Pulkovo observatory and SL developed in the Paris observatory show similar FCN amplitude and phase
variations, and hence they are nearly equivalent from this point if view.
However, it is not the case for CPO modelling.
Using different models for space geodesy data processing can lead to substantial biases and systematic differences in results, as was shown,
for example, for VLBI UT1 Intensives.

Unfortunately, there is no generally accepted option to account for the CPO in various applications, in particular in space geodesy data processing.
The SL model recommended by the IERS Conventions (McCarthy \& Petit 2004) is not an CPO model and can compensate only about a half of the full effect.
It seems advisable that IERS consider another model more close to the actual CIP motion to recommend it in the Conventions.
A new model should also provide accurate CPO prediction for several weeks.

The simplest way to eliminate the main deficiencies of the IERS (SL) model would be to include bias removed during the model construction (Lambert 2009).
However, it still is perhaps too smoothed.
Besides it is updated only once a year (Lambert, personal communication).

NEOS model show much better approximation of the VLBI data, but has poor prediction (Malkin 2010).

The ZM2 model seems to be preferable as providing the best approximation of the VLBI results and CPO prediction.

\smallskip
{\it Acknowledgements.} The author is grateful to the organizers of the conference for the travel support.

\smallskip

\vspace*{0.7cm}

\noindent {\large 5. REFERENCES}

{

\leftskip=5mm
\parindent=-5mm

\smallskip

B\"ockmann, S., Artz, T., Nothnagel, A., Tesmer, V. International VLBI Service for Geodesy and Astrometry: Earth orientation
parameter combination methodology and quality of the combined products. 2010, \jgr, 115(B4), B04404.

Gubanov, V.S., 2010, ``New estimates of retrograde free core nutation parameters'', Astron. Letters, 36(6), pp.~444--451.

Herring, T.A., Mathews, P.M., Buffett, B.A., 2002, ``Modelling of nutation-precession: Very long baseline interferometry results'',
\jgr, 107(B4), pp.~2069--2080.

IERS Annual Report 2007, 2009, Verlag des Bundesamts f\"ur Kartographie und Geod\"asie, Frankfurt am Main.

Lambert, S.B., 2009, ``Empirical Model of the Free Core Nutation (Technical Note)'',\\
http://syrte.obspm.fr/{$\sim$}lambert/fcn/notice.pdf

Malkin, Z.M., 2007, ``Empiric Models of the Earth's Free Core Nutation'', Solar System Research, 41(6), pp.~492--497.

Malkin, Z.M., 2010, ``Analysis of the Accuracy of Prediction of the Celestial Pole Motion'', Astronomy Reports, 54(11), pp.~1053-1061.

McCarthy, D.D., Petit, G. (eds), 2004, ``IERS Conventions (2003)'', IERS Technical Note 32,
Verlag des Bundesamts f\"ur Kartographie und Geod\"asie, Frankfurt am Main.

Shirai, T., Fukushima, T., Malkin Z., 2005, ``Detection of phase disturbances of free core nutation of the Earth and their concurrence
with geomagnetic jerks'', Earth Planets Space, 57(2), pp.~151--155.

}

\end{document}